\documentclass[aps,prb,twocolumn,superscriptaddress,showpacs]{revtex4}
\usepackage{graphicx}
\begin{document}

\title{Observation of an inter-sublattice exchange magnon in CoCr$_2$O$_4$ \\
and analysis of magnetic ordering}

\author{D. Kamenskyi}
\affiliation{Dresden High Magnetic Field Laboratory (HLD),
Helmholtz-Zentrum Dresden-Rossendorf, 01314 Dresden, Germany}

\author{H. Engelkamp}
\affiliation{High Field Magnet Laboratory, Institute for Molecules
and Materials, Radboud University Nijmegen, 6525 ED Nijmegen, The
Netherlands}

\author{T. Fischer}
\affiliation{Dresden High Magnetic Field Laboratory (HLD),
Helmholtz-Zentrum Dresden-Rossendorf, 01314 Dresden, Germany}

\author{M. Uhlarz}
\affiliation{Dresden High Magnetic Field Laboratory (HLD),
Helmholtz-Zentrum Dresden-Rossendorf, 01314 Dresden, Germany}

\author{J. Wosnitza}
\affiliation{Dresden High Magnetic Field Laboratory (HLD),
Helmholtz-Zentrum Dresden-Rossendorf, 01314 Dresden, Germany}

\author{B. P. Gorshunov}
\affiliation{A. M. Prokhorov Institute of General Physics, Russian
Academy of Sciences, 119991 Moscow, Russia} \affiliation{Moscow
Institute of Physics and Technology (State University), 141700
Dolgoprudny, Moscow Region, Russia} \affiliation{1. Physikalisches
Institut, Universit\"{a}t Stuttgart, Pfaffenwaldring 57, 70550
Stuttgart, Germany}

\author{\\ G. A. Komandin}
\affiliation{A. M. Prokhorov Institute of General Physics, Russian
Academy of Sciences, 119991 Moscow, Russia}

\author{A. S. Prokhorov}
\affiliation{A. M. Prokhorov Institute of General Physics, Russian
Academy of Sciences, 119991 Moscow, Russia} \affiliation{Moscow
Institute of Physics and Technology (State University), 141700
Dolgoprudny, Moscow Region, Russia}

\author{M. Dressel}
\affiliation{1. Physikalisches Institut, Universit\"{a}t Stuttgart,
Pfaffenwaldring 57, 70550 Stuttgart, Germany}

\author{A. A. Bush}
\affiliation{Moscow State Institute of Radio-Engineering,
Electronics, and Automation (Technical University), 117464 Moscow,
Russia}

\author{V. I. Torgashev}
\affiliation{Faculty of Physics, Southern Federal University, 344090
Rostov-on-Don, Russia}

\author{A. V. Pronin} \email{a.pronin@hzdr.de}
\affiliation{Dresden High Magnetic Field Laboratory (HLD),
Helmholtz-Zentrum Dresden-Rossendorf, 01314 Dresden, Germany}

\date{\today}

\begin{abstract}

We report on an investigation of optical properties of multiferroic
CoCr$_{2}$O$_{4}$ at terahertz frequencies in magnetic fields up to
30 T. Below the ferrimagnetic transition (94 K), the terahertz
response of CoCr$_{2}$O$_{4}$ is dominated by a magnon mode, which
shows a steep magnetic-field dependence. We ascribe this mode to an
exchange resonance between two magnetic sublattices with different
$g$-factors. In the framework of a simple two-sublattice model (the
sublattices are formed by Co$^{2+}$ and Cr$^{3+}$ ions), we find the
inter-sublattice coupling constant, $\lambda = - (18 \pm 1)$ K, and
trace the magnetization for each sublattice as a function of field.
We show that the Curie temperature of the Cr$^{3+}$ sublattice,
$\Theta_{2}$ = $(49 \pm 2)$ K, coincides with the temperature range,
where anomalies of the dielectric and magnetic properties of
CoCr$_{2}$O$_{4}$ have been reported in literature.

\end{abstract}


\maketitle

CoCr$_{2}$O$_{4}$ is a ferrimagnetic spinel compound with a complex
network of competing magnetic interactions. \cite{menyuk, tomiyasu}
Both, Co$^{2+}$ (A sites of the spinel structure) and Cr$^{3+}$ ions
(B1 and B2 sites), are magnetic. Below the Curie temperature, $T_{C}
= 94$ K, the system exhibits a long-range ferrimagnetic order.
\cite{menyuk} At $T_{S} = 26$ K, a structural transition occurs.
Below this temperature, an incommensurate conical (i.e. uniform plus
transverse spiral) magnetic structure sets in. At $T_{\rm lock-in} =
15$ K, the magnetic structure becomes commensurate -- the period of
the spin spiral ``locks" to the lattice parameter. \cite{tomiyasu}

The spiral order most likely survives above $T_{S}$, but on short
ranges only. \cite{menyuk, tomiyasu} At $T_{\rm kink} = 50$ K,
anomalies in dielectric and magnetic properties of CoCr$_{2}$O$_{4}$
have been reported \cite{pronin, lawes} and tentatively attributed
to the formation of this incommensurate short-range spiral magnetic
order. \cite{lawes}

In 2006, Yamasaki \textit{et al.} have discovered the structural
transition at $T_{S} = 26$ K to be accompanied by the emergence of a
spontaneous electric polarization, which direction can be reversed
by applying a magnetic field. \cite{yamasaki} As multiferroics are
appealing because of basic physical interest as well as their
potential technological applications, \cite{kimura} the reported
multiferroicity of CoCr$_{2}$O$_{4}$ has triggered a number of
experimental studies of the compound. \cite{lawes, choi, mufti,
chang, torgashev} To date, there is, however, no optical data
reported at terahertz frequencies. This is the frequency region
where \textit{e.g.} electromagnons have been observed in some other
multiferroic systems. \cite{pimenov, sushkov}

Here, we report on an optical study of CoCr$_{2}$O$_{4}$ at
terahertz (or far-infrared) frequencies in magnetic fields up to 30
T. We do not find any evidence for electromagnons. However, below
$T_{C}$, we observe a resonance mode, which is highly sensitive to
external magnetic field. We ascribe this mode to a ferrimagnetic
inter-sublattice exchange resonance, originally considered
theoretically by Kaplan and Kittel. \cite{kaplan} Applying this
model to CoCr$_{2}$O$_{4}$ allows us to separate the contributions
of the Co$^{2+}$ and Cr$^{3+}$ sublattices to the total
magnetization, to extract the magnetization for each sublattice as a
function of field, and to find the inter-sublattice coupling
constant. Furthermore, we show that the Curie temperature of the
Cr$^{3+}$ sublattice coincides with $T_{\rm kink} = 50$ K. Thus, the
onset of the short-range spiral component must be related to the
ordering in the Cr$^{3+}$ sublattice.

The investigated CoCr$_{2}$O$_{4}$ samples have been synthesized
from Co$_{3}$O$_{4}$ and Cr$_{2}$O$_{3}$ powders. CoCr$_{2}$O$_{4}$
powder has been pressed into pellets of 10 mm in diameter and 1 mm
in thickness. X-ray diffraction measurements prove the cubic
symmetry (space group $Fd\bar{3}m$) with no indication of spurious
phases. The lattice constant, $a = 8.328(2)$ {\AA}, is in good
agreement with published results. \cite{mansour, casado} More
details on sample preparation, characterization, and thermodynamic
properties of CoCr$_{2}$O$_{4}$ are given in our recent works.
\cite{pronin, uhlarz}

The optical measurements have been made in transmission mode, using
two different setups. The first setup was a spectrometer equipped
with backward-wave oscillators (BWOs) as sources of coherent and
frequency-tunable radiation. \cite{kozlov, gorshunov} The
measurements were done at (sub)terahertz frequencies ($\nu =
\omega/2\pi$ = $8.5 \div 41$ cm$^{-1}$ = $250 \div 1230$ GHz = $1
\div 5$ meV) and at temperatures between 4 and 300 K. The radiation
was linearly polarized. The absolute values of optical transmission
(normalized to the empty-channel measurements) were obtained in
essentially the same way as described, e.g., in Ref.
\onlinecite{mukhin}.

A commercial split-coil magnet, embedded into an optical cryostat,
was utilized for measurements in magnetic fields up to 8 T. In these
measurements, we used three different geometries: $\mathbf{k} \perp
\mathbf{H} \parallel \tilde{\mathbf{h}}$, $\mathbf{k} \perp
\mathbf{H} \perp \tilde{\mathbf{h}}$, and $\mathbf{k} \parallel
\mathbf{H} \perp \tilde{\mathbf{h}}$ (here $\mathbf{H}$ is the
external static magnetic field; $\mathbf{k}$ and
$\tilde{\mathbf{h}}$ are the wave vector and the magnetic component
of the probing electromagnetic radiation, respectively).

In the first geometry ($\mathbf{k} \perp \mathbf{H} \parallel
\tilde{\mathbf{h}}$), where electromagnons, if they exist, could be
excited, we did not observe any resonance absorption lines.
Furthermore, these spectra show no detectable changes induced by
applying magnetic fields up to 8 T.

\begin{figure}[t]\vspace{0cm}
\centering
\includegraphics[width=7 cm,clip]{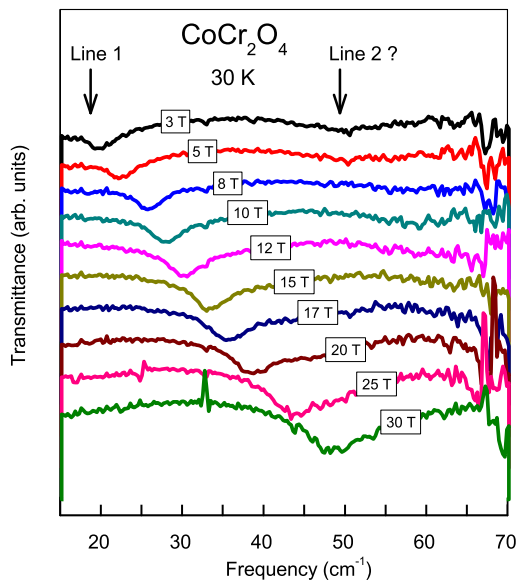}
\vspace{0cm} \caption{(Color online) Examples of raw far-infrared
transmittance spectra of CoCr$_{2}$O$_{4}$ in Faraday geometry. The
absorption line, marked as ``Line 1", is discussed in the course of
the article. At the higher-frequency end of the spectra, there is an
indication of another, very broad, line (``Line 2 ?").}
\label{magnon0}
\end{figure}

In the two other geometries (where $\mathbf{H} \perp
\tilde{\mathbf{h}}$), we observed absorption lines. We did not
detect any differences between the results for these two geometries,
although the data obtained for $\mathbf{k}\parallel \mathbf{H}$
(Faraday geometry) are somewhat less noisy due to peculiarities of
the split-coil magnet design. Thus, in the following we discuss the
BWO data collected in the Faraday geometry.

The second optical setup was a commercial Fourier-transform infrared
(FTIR) spectrometer (Bruker IFS113v) combined with a
continuous-field 33-Tesla Bitter magnet, at the High Field Magnet
Laboratory in Nijmegen. \cite{wiegers} The measurements were
performed in the Faraday geometry ($\mathbf{k}\parallel \mathbf{H}
\perp \tilde{\mathbf{h}}$) at 4 and 30 K. A mercury lamp was used as
a radiation source. The far-infrared radiation was detected using a
custom-made silicon bolometer operating at 1.4 K. For both
temperatures, the FTIR spectra were measured in 0, 3, 5, 8, 10, 12,
15, 17, 20, 25, and 30 T. The optical data were collected between 15
and 70 cm$^{-1}$ ($450 \div 2100$ GHz, $1.9 \div 8.7$ meV), using a
200-$\mu$m mylar beamsplitter and a scanning velocity of 50 kHz. At
each field, at least 100 scans were averaged.

\begin{figure}[t]\vspace{0cm}
\centering
\includegraphics[width=7 cm,clip]{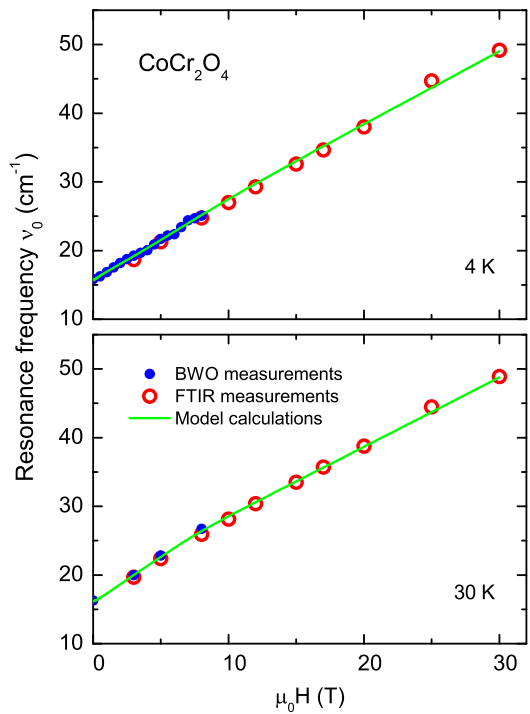}
\vspace{0cm} \caption{(Color online) Field dependence of the magnon
frequency for 4 and 30 K. Symbols are the experimental data, lines
represent the results of simultaneous fits of the optical data with
Eq. \ref{exchange} and of magnetization data from Ref.
\onlinecite{pronin} with Eqs. \ref{mag} and \ref{field}. Note that
the onset of the vertical scales is at 10 cm$^{-1}$.}
\label{magnon3}
\end{figure}

In the transmission spectra, obtained by use of both setups, we
reliably observed an intensive absorption line, which is highly
sensitive to the applied magnetic field and temperature (``Line 1"
in Fig. \ref{magnon0}). At the higher-frequency end of the FTIR
spectra, we possibly observe another, broad and week, absorption
line (``Line 2 ?" in Fig. \ref{magnon0}). The position of this line
shifts to higher frequencies (and thus out of the
measurement-frequency window) in applied magnetic field. We
tentatively attribute this line to an antiferromagnetic resonance
within the Cr$^{2+}$ sublattice. Hereafter, we solely discuss the
``Line 1".

The field dependence of the frequency of ``Line 1", $\nu_{0}$, is
shown in Fig. \ref{magnon3}. The resonance frequency is basically
proportional to the applied static field with a zero-field gap. At 4
K, the gap is 15.8 cm$^{-1}$ (475 GHz, 1.97 meV), and at 30 K, it is
16.3 cm$^{-1}$ (489 GHz, 2.02 meV). At low fields [$\mu_{0}H
\lesssim 10$ T], the dimensionless slopes of the $\nu_{0} (H)$
curves, $h\nu_{0}/\mu_{B}\mu_{0}H$, reach 2.5 (at 4 K) and 2.8 (at
30 K). Let us note, that the typical slopes of the $\nu_{0} (H)$
curves for ferromagnetic resonances in the Co$^{2+}$- or
Cr$^{3+}$-based compounds are determined by the gyromagnetic ratios
of the Co$^{2+}$ and Cr$^{3+}$ ions and amount to 2.2 and 1.95,
respectively. \cite{altshuler}

Resonance modes in ferrimagnetic substances have originally been
considered theoretically by Kaplan and Kittel. \cite{kaplan} For a
system with two magnetic sublattices, they predicted the existence
of two magnetic-resonance lines. The first of them is associated
with the conventional spin precession. In CoCr$_{2}$O$_{4}$, such
mode has been observed by electron-spin-resonance studies at
frequencies below 100 GHz in fields up to 10 T. \cite{stickler,
funahashi} The second mode, the inter-sublattice exchange resonance,
is supposed to have a large zero-field gap defined by the exchange
interaction between the sublattices, to harden in an applied
magnetic field, and to have a steep field dependence. \cite{kaplan2}
Our observed mode demonstrates this very behavior (Fig.
\ref{magnon3}). Thus, it is reasonable to associate it with the
Kaplan-Kittel inter-sublattice exchange resonance.

In the following, we show that using a simple two-sublattice
Kaplan-Kittel model, we can consistently describe our data on the
magnon frequency and our earlier data on magnetization (Ref.
\onlinecite{pronin}). From this description, we obtain the magnetic
moments of the sublattices and the inter-sublattice exchange
constant.

In our model, we consider two effective magnetic sublattices.
Coupling within the first sublattice is provided by exchange
interactions between Co$^{2+}$ and Cr$^{3+}$ ions, $J_{AB}$. At
$\Theta_{1}$ = $T_{C}$ = 94 K, this sublattice orders. Important is
that only non-collinearly ordered spins of the Co$^{2+}$ ions
contribute to the magnetic moment, while the spins of the Cr$^{3+}$
ions do not. This is because of geometrical frustration of the
interaction between the Cr$^{3+}$ ions, $J_{BB}$. \cite{tomiyasu}
Thus, one can consider the first sublattice to be formed by
Co$^{2+}$ ions only.

When temperature decreases further, the Cr$^{3+}$ spins start to
contribute (negatively) to the net magnetization. Thus, as the
second sublattice, one can consider the sublattice formed by
Cr$^{3+}$ ions. Let us note, that because of geometrical
frustration, the effective coupling within the Cr$^{3+}$ sublattice
is significantly smaller than $J_{BB}$. The Curie temperature of
this sublattice, $\Theta_{2}$, is somewhere below 94 K. This
temperature will be one of our fit parameters (we take $\Theta_{2}$
to be field independent).

For a two-sublattice ferrimagnet, the net (measurable) magnetization
is the sum of the magnetizations of the sublattices, $M(T, H) =
M_{1}(T, H) + M_{2}(T, H)$. $M_{1}$ and $M_{2}$ have opposite signs;
the inter-sublattice exchange constant $\lambda$ is negative. Within
the molecular-field approximation, the reduced magnetization of each
sublattice, $y_i \equiv M_i/M_i^0$ [here and thereafter $i = \{1,2\}
\equiv \{\textrm{Co}^{2+}, \textrm{Cr}^{3+}\}$ and $M_{i}^{0}$ is
the zero-temperature magnetization of the $i$-th sublattice], is
given by the Brillouin function: \cite{vonsovsky, kuzmin}
\begin{equation} \label{mag}
y_i = B_{S_{i}}(x_{i}) \equiv \frac{2S_{i}+1}{2S_{i}} \coth
\frac{2S_{i}+1}{2S_{i}} x_{i} - \frac{1}{2S_{i}} \coth
\frac{x_{i}}{2S_{i}},
\end{equation}
where $S_{i}$ is the spin magnetic moment for ions of the $i$-th
sublattice and $x_{i}$ is defined as:
\begin{eqnarray}  \label{field}
x_{i} = \frac{\mu_{B}} {k_{B}T} \left[\frac{3S_{i}}{S_{i}+1} \left(
\frac{k_{B} \Theta_{i} M_{i}} {\mu_{B} M_{i}^{0}} + \frac{k_{B}
\lambda  M_{j}} {\mu_{B} M_{i}^{0}} \right)+ H M_{i}^{0}  \right],
\\i \neq j. \nonumber
\end{eqnarray}
In our case, Co$^{2+}$ and Cr$^{3+}$ ions have equal spin moments,
$S_{1} = S_{2} = 3/2$.

\begin{figure}[t]\vspace{0cm}
\centering
\includegraphics[width=7 cm,clip]{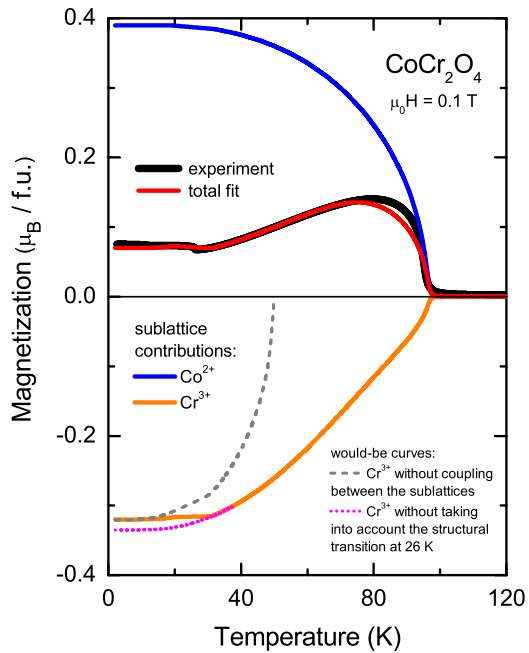}
\vspace{0cm} \caption{(Color online) Example of a fit by use of Eqs.
\ref{mag} and \ref{field} to the magnetization data obtained at 0.1
T [simultaneously, the magnon-frequency data were fitted with the
same set of free parameters, see Fig. \ref{magnon3}]. Contributions
of the Co$^{2+}$ and Cr$^{3+}$ sublattices are shown with opposite
signs. The contribution of the Cr$^{3+}$ sublattice diminishes at
$T_{C}$, rather than at $\Theta_{2}$ = 49 K, because of the coupling
between the sublattices.} \label{magnon4}
\end{figure}

For the resonance frequency $\nu_{0}$, we use a modified
Kaplan-Kittel equation: \cite{kaplan, torgashev}
\begin{eqnarray}
\left( \frac{h \nu_{0}}{\mu_{B} g_{1}} + \frac{k_{B}}{\mu_{B}}
\lambda M_{2} + H \right)\times \label{exchange} \\
\left( \frac{h \nu_{0}}{\mu_{B} g_{2}} - \frac{k_{B}}{\mu_{B}}
\lambda M_{1} + H \right) + \left(\frac{k_{B}}{\mu_{B}} \lambda
\right)^{2} M_{1}M_{2} = 0, \nonumber
\end{eqnarray}
where $g_{1}$ = 2.2 and $g_{2}$ = 1.95 are the gyromagnetic ratios
for the Co$^{2+}$ and Cr$^{3+}$ ions, respectively. \cite{altshuler}
In Eqs. \ref{field} and \ref{exchange}, magnetization is in
dimensionless units. In our model, we neglect anisotropy, as our
measurements have been performed on powder samples. \cite{powder}

Unlike in collinear ferrimagnetics, in CoCr$_{2}$O$_{4}$ the vector
of the magnetization of each sublattice in an external magnetic
field can change not only its orientation, but also its size. In our
calculations, we take this into account. The corresponding increase
in $M^{0}_{1}$ and $M^{0}_{2}$ is supposed to persist until the
spin-only values for Co$^{2+}$ and Cr$^{3+}$ ions of 3 $\mu_{B}$/ion
are reached. Recent experiments in pulsed magnetic fields have shown
that even in fields of up to 60 T, there are no signs of saturation
in the magnetization. \cite{tsurkan}

Using Eqs. \ref{mag} -- \ref{exchange}, we simultaneously fit the
magnetization data, obtained in Ref. \onlinecite{pronin}, and our
results for the resonance frequency. \cite{30T} Results of these
fits are shown in Figs. \ref{magnon3} and \ref{magnon4} together
with experimental data. We find that the best description of all our
experimental data can be reached with $\lambda = - (18 \pm 1)$ K and
$\Theta_{2}$ = $(49 \pm 2)$ K.

The value found for the ordering temperature of the Cr$^{3+}$
sublattice, $\Theta_{2}$, coincides with $T_{\rm kink}$ -- the
temperature, which is possibly related to the formation of the
incommensurate short-range spiral magnetic order. \cite{lawes} Our
findings show that this short-range order appears to be due to the
involvement of the Cr$^{3+}$ ions. The relatively large value of the
coupling constant, $\lambda = - 18 $ K, may explain why the $T_{\rm
kink}$-related features appear to be broad. \cite{tomiyasu, lawes}

\begin{figure}[]\vspace{0cm}
\centering
\includegraphics[width=7 cm,clip]{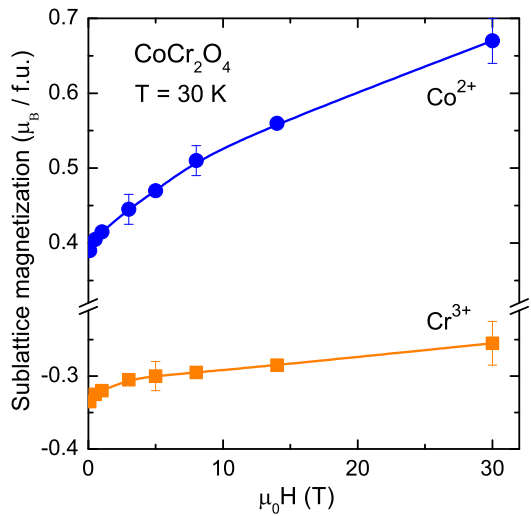}
\vspace{0cm} \caption{(Color online) Magnetization of the Co$^{2+}$
and Cr$^{3+}$ sublattices in CoCr$_{2}$O$_{4}$, as obtained from the
model fits with Eqs. \ref{mag} -- \ref{exchange}. The lowest-field
points are at 0.1, 0.5, and 1 T.} \label{magnon5}
\end{figure}

Figure \ref{magnon5} shows the magnetic-field dependence of the
calculated sublattice magnetizations. We show here the results
obtained at $T = 30$ K, i.e. just above the structural transition.
As it can be seen from Fig. \ref{magnon4}, the lower-temperature
results differ only marginally from the shown in Fig. \ref{magnon5}.
We have found that within our model, the observed behavior of the
experimental data at $T_{S}$ can be best described by a slight
decrease in the absolute value of $M_{Cr}$ at $T_{S}$ (the
$g$-factors of the Co$^{2+}$ and Cr$^{3+}$ ions are presumed to be
independent of temperature).

The contribution of each sublattice to the total magnetic moment in
zero field can be estimated from Fig. \ref{magnon5}. We find
$M_{Co}$(0~T) = 0.4 $\mu_{B}$/f.u. and $M_{Cr}$(0~T) = -- 0.33
$\mu_{B}$/f.u. We note, that although in fields smaller than the
coercive force (0.3 T, Ref. \onlinecite{pronin}) the $M_{i}$ values
are influenced by the magnetizing/demagnetizing processes, this
influence hardly affects the above result [in Fig. \ref{magnon5},
only the lowest-field point (0.1 T) is below the coercive-force
value, the next points are at 0.5 and 1 T].

The proposed model gives a natural explanation for the steep slope
of the $\nu_{0} (H)$ curves [Fig. \ref{magnon3}]. As can be seen
from Eq. \ref{exchange}, this is due to the increase of $M_{i}$ as a
function of field. Similarly, the temperature evolution of $\nu_{0}
(H)$ is explained as being due to the temperature evolution of the
$M_{i} (H)$ curves.

Summarizing, the far-infrared optical response of CoCr$_{2}$O$_{4}$
below the Curie temperature, $T_{C} = 94$ K, is dominated by a
magnon mode. Ascribing the magnon to an inter-sublattice exchange
resonance (the sublattices are formed by the Co$^{2+}$ and Cr$^{3+}$
ions) and applying a modified Kaplan-Kittel model of two interacting
sublattices allows us to consistently describe our experimental data
on the magnon frequency and earlier magnetization measurements, to
obtain the magnetization for each of the sublattices, and to find
the inter-sublattice exchange constant, $\lambda$ = -- 18 K. The
Curie temperature of the Cr$^{3+}$ sublattice is found to coincide
with the temperature, where the incommensurate short-range spiral
magnetic order is believed to set in, $T_{\rm kink} = 50$ K.

We thank M. D. Kuz'min for useful discussions and Papori Gogoi for
assistance in measurements. Parts of this work were supported by
EuroMagNET II (EU contract No. 228043), by the Russian Foundation
for Basic Research (Grant No. 12-02-00151) and by the Ministry of
Education and Science of the Russian Federation (contract No.
14.A18.21.0740).


\begin{thebibliography}{99}
\bibitem{powder} In practice, this approximation would also be valid for single
crystals, as the magnitude of the anisotropy field in
CoCr$_{2}$O$_{4}$ is below 0.1 T [Ref. \onlinecite{funahashi}], i.e.
significantly smaller than all terms in Eqs. \ref{field} and
\ref{exchange} for external fields above 0.1 T.
\bibitem{30T} Since no experimental data on the temperature dependence of
the magnetization at 30 T are available, we used a linear
extrapolation of the magnetization in accordance with the $M(H)$
measurements of Ref. \onlinecite{tsurkan}.
\end{thebibliography}
\end{document}